\documentclass[12pt]{spieman}  
\usepackage{anyfontsize} 
\usepackage{titlesec}
\usepackage{mathrsfs}
\usepackage{amsmath,amsfonts,amssymb}
\usepackage{graphicx}
\usepackage{setspace}
\usepackage{tocloft}
\usepackage{svg}
\usepackage{hyperref}
\usepackage{float}
\usepackage{dirtytalk}
\usepackage[margin=1in]{geometry}
\usepackage{lmodern}
\usepackage{comment}
\usepackage{booktabs}
\usepackage[nameinlink,noabbrev,capitalize]{cleveref}
\usepackage[mathlines]{lineno}
\usepackage{siunitx}
\usepackage{parskip}
\usepackage[title,toc,page]{appendix}
\usepackage{xr}

\usepackage{nameref}

\newcommand{\supref}[1]{\hyperref[#1]{Supplementary Material~\ref*{#1}}}
\newcommand{\suptableref}[1]{\hyperref[#1]{Supplementary~Table~\ref*{#1}}}
\newcommand{\supfigureref}[1]{\hyperref[#1]{Supplementary~Figure~\ref*{#1}}}

\newcommand{\minifinder}{\texttt{IDT}}
\newcommand{\aqua}{\texttt{AQuA}}
\newcommand{\astrobeats}{\texttt{Astro-BEATS}}
\newcommand{\deltaF}{\ensuremath{\Delta F/F_0}}

\title{Ca$^{2+}$ transient detection and segmentation with the Astronomically motivated algorithm for Background Estimation And Transient Segmentation (Astro-BEATS)}

\author[a,b]{Bolin Fan}
\author[c,d]{Anthony Bilodeau}
\author[c,d]{Frédéric Beaupré}
\author[c,d]{Theresa Wiesner}
\author[d,e,f]{Christian Gagné}
\author[c,d,g]{Flavie Lavoie-Cardinal}
\author[a,b,*]{Renée Hlo\v{z}ek}
\affil[a]{Dunlap Institute for Astronomy and Astrophysics, University of Toronto, Toronto, Canada}
\affil[b]{David A. Dunlap Department for Astronomy and Astrophysics, University of Toronto, Toronto, Canada}
\affil[c]{CERVO Brain Research Center, Québec, Canada}
\affil[d]{Institute Intelligence and Data, Université Laval, Québec, Canada}
\affil[e]{Department of Electrical Engineering and Computer Engineering, Université Laval, Québec, Canada}
\affil[f]{Canada CIFAR AI Chair, affiliated to Mila}
\affil[g]{Department of Psychiatry and Neuroscience, Université Laval, Québec, Canada}

\cftpagenumbersoff{figure}
\cftpagenumbersoff{table}

\begin{document}

\maketitle

\begin{abstract}
\section*{Significance}
Fluorescence-based Ca$^{2+}$-imaging is a powerful tool for studying localized neuronal activity, including miniature Synaptic Calcium Transients, providing real-time insights into synaptic activity. These transients induce only subtle changes in the fluorescence signal, often barely above baseline, which poses a significant challenge for automated synaptic transient detection and segmentation.
\section*{Aim}
Detecting astronomical transients similarly requires efficient algorithms that will remain robust over a large field of view with varying noise properties. We leverage techniques used in astronomical transient detection for miniature Synaptic Calcium Transient detection in fluorescence microscopy.
\section*{Approach}
We present \astrobeats, an automatic miniature Synaptic Calcium Transient segmentation algorithm that incorporates image estimation and source-finding techniques used in astronomy and designed for Ca$^{2+}$-imaging videos. \astrobeats~uses the Rolling Hough Transform filament detector to construct an estimate of the expected (transient-free) fluorescence signal of both the dendritic foreground and the background. Subtracting this baseline signal yields difference images displaying transient signals. We use Density-Based Spatial Clustering of Applications with Noise to find sources clustered in spatial and temporal space.
\section*{Results}
\astrobeats~outperforms current threshold-based approaches for synaptic Ca$^{2+}$ transient detection and segmentation. The produced segmentation masks can be used to train a supervised deep learning algorithm for improved synaptic Ca$^{2+}$ transient  detection in Ca$^{2+}$-imaging data.  The speed of \astrobeats\ and its applicability to previously unseen datasets without re-optimization makes it particularly useful for generating training datasets for deep learning-based approaches.
\section*{Conclusion}
\astrobeats~greatly reduces the time needed for the annotation of synaptic Ca$^{2+}$ transient and removes the significant overhead of human expert annotation, enabling consistent analysis of new Ca$^{2+}$-imaging datasets.

\end{abstract}
{\noindent \footnotesize\textbf{*}Address all correspondence to Renée Hlo\v{z}ek at hlozek@dunlap.utoronto.ca}


\begin{spacing}{2}   

\section{Introduction} \label{sec:intro}
Detecting transient and dynamic events within biological systems is essential to deepen our understanding of complex cellular processes. A variety of fluorescence microscopy techniques, including Ca$^{2+}$-imaging, have been developed to capture these events~\cite{kerr2008imaging, grienberger2012imaging,cuny2022microscopy}. When combined with sensitive Ca$^{2+}$ indicators such as GCaMP6\cite{gcamp}, fluorescence microscopy provides sufficient contrast to resolve subcellular fluctuations in Ca$^{2+}$ concentration linked to local neuronal activity and miniature Synaptic Calcium Transients (mSCTs)\cite{andreae2015spontaneous, murphy1995mapping, reese2016single, kavalali2015mechanisms, beaupre2024quantitative}. In time-lapse Ca$^{2+}$-imaging,  mSCTs are characterized by small ($\mathscr{O}(1)-\mathscr{O}(10) \mu$m$^2$) and short (ms-s) fluctuations in the fluorescence signal of the Ca$^{2+}$-sensor\cite{beaupre2024quantitative,murphy1995mapping,andreae2015spontaneous,murthy2000dynamics,reese2016single}. Extracting these transient signals from microscopy data remains a major analytical challenge, as they must be reliably distinguished from background noise and other sources of brightness variability\cite{murphy1995mapping,beaupre2024quantitative,yu2024imaging}. 

Although threshold-based transient detection can be effectively find mSCTs when fixed regions of interest can be detected\cite{agarwal2017transient,srinivasan2015ca2}, these methods struggle to detect events that evolve spatially and temporally (i.e., spread out from their point of origin)\cite{beaupre2024quantitative}. Manual annotation of mSCTs is both highly challenging and time-consuming, and is not scalable to large imaging Ca$^{2+}$ datasets\cite{beaupre2024quantitative,vohs2008manual}. Deep learning (DL) strategies have been proposed for detecting sparse, low-intensity transients but require annotated datasets for training\cite{beaupre2024quantitative,dotti2024deep,xu2019u}. Generating these training datasets can be challenging, especially when manual annotation of the dataset is not possible due to the complexity of the task or the size of the dataset. Using a publicly available Ca$^{2+}$-imaging dataset of mSCTs\cite{beaupre_2026_18561819} (\autoref{sec:dataset}), we leverage the similarity between astronomical transients and mSCTs to utilize tools from astronomy to perform mSCT detection and segmentation.

\begin{figure}
\begin{center}
\begin{tabular}{c}
\includegraphics[width=0.85\textwidth]{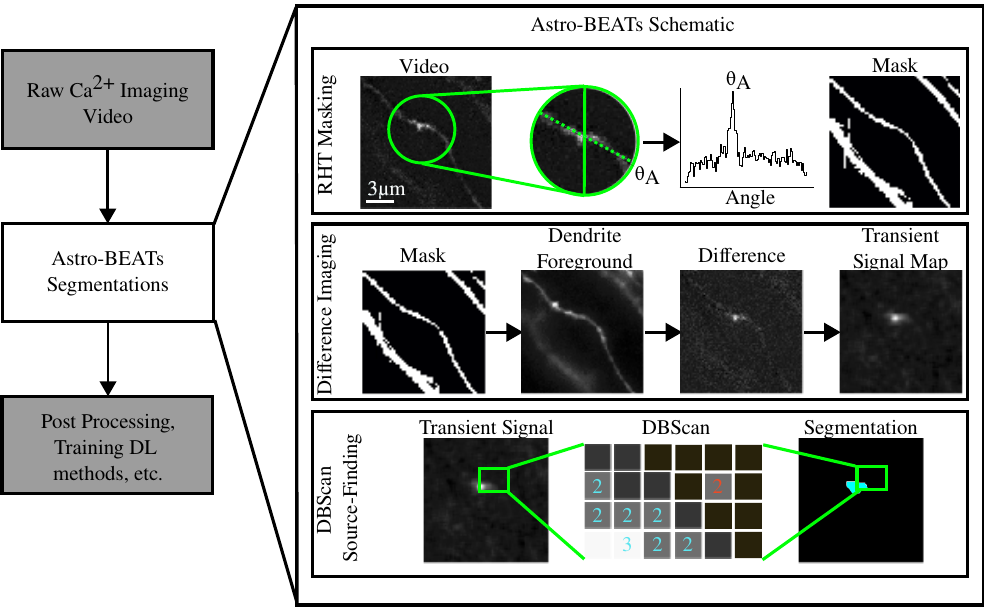}
\end{tabular}
\end{center}
\caption{\label{fig:figure_1_intro_schematic} Overview of \astrobeats~in the context of mSCTs analysis. Raw Ca$^{2+}$-imaging videos are processed using \astrobeats, an unsupervised source-finding algorithm used to generate segmentation masks of transients in these videos. The RHT\cite{rht_clark} is used to generate a dendritic foreground mask, a mask of pixels that contain the foreground GCaMP fluorescence signal (top row, described in \supref{sec:rht}). A dynamic estimate of the GCaMP6f fluorescence signal in the dendrites is used to estimate the average foreground in each frame. We take the difference image between this weighted foreground and our raw Ca$^{2+}$ video to isolate mSCTs from the foreground (middle row, described in \supref{sec:decompose}). Finally, density-based spatial clustering of applications with noise (DBSCAN) source-finding is used on pixels in the difference image to identify and segment transients (bottom row, described in \autoref{sec:astrobeats_sources}).}
\end{figure}

The need in astronomy to quickly and automatically detect and classify objects has led to the development of several image-analysis techniques to separate rapidly-varying astrophysical signals from averaged (transient-free) images of the sky\cite{zackay2016proper,masias2012review}. Detecting and identifying these new sources of light in the sky before they fade from view is crucial to be able to study the origin, properties, and emission mechanisms of these objects\cite{zackay2016proper}. Astronomical transients can occur at any place in an astronomical image, in a way analogous to mSCTs in a Ca$^{2+}$-imaging videos. Transient detection proceeds through a process called \emph{difference imaging}, where an image of the sky at a particular time (when a transient might have occurred) is compared to an expected \emph{background image} of the same region of the sky at an earlier time, produced by averaging multiple earlier observations\cite{td_1,zackay2016proper}. The difference between an image taken at the particular time and the slowly-changing, averaged background yields an image that contains mostly random brightness fluctuations but also contains a bright point at the transient source sky location. Source-finding techniques then classify the bright points in the difference image as objects of interest and separate them from sources of noise. Due to the large volumes of data in astronomy, difference imaging\cite{td_1,td_2} and source-finding algorithms\cite{blobcat2012,hancock2012} have evolved to be automated and efficient with little to no human input through manual annotation\cite{big_data}. Astronomical transient detection algorithms are designed to solve a problem similar to those related to mSCT detection. We leverage the similarities between the data types and present the Astronomically-motivated algorithm for Background Estimation And Transient Segmentation (\astrobeats), a fully automated detection and segmentation algorithm of mSCTs in Ca$^{2+}$-imaging datasets. We use it to generate segmentation masks of comparable quality to manual expert annotations. Transients detected by \astrobeats~can then be analyzed directly, or used as an annotated training dataset for DL models (see \autoref{fig:figure_1_intro_schematic}).

\section{Methods} \label{sec:methods}
\subsection{Dataset}\label{sec:dataset}
To evaluate the performance of the algorithm, we apply \astrobeats~on a publicly available Ca$^{2+}$-imaging dataset\cite{beaupre_2026_18561819}. The sample preparation and Ca$^{2+}$-imaging video acquisition techniques used to generate the dataset are described in detail in Ref.~\citenum{beaupre2024quantitative}. In brief, the dataset consists of Ca$^{2+}$-imaging videos from cultured hippocampal neurons expressing GCaMP6f. Ca$^{2+}$ recordings are acquired using Total Internal Reflection Fluorescence Microscopy (TIRF) with oblique illumination. Each frame of the Ca$^{2+}$-imaging video has a size of $512\times512$ pixels with a pixel size of $0.16~\mu\mathrm{m}\times0.16~ \mu\mathrm{m}$. The videos are acquired with a 10 Hz (10 frames/s) sampling frequency for 60 to 300 seconds. The pixel values are defined as $\Delta F / F_0 = \frac{F - F_0}{F_0}$, where $F$ is the fluorescence signal recorded by the microscope and $F_0$ is a baseline fluorescence intensity obtained during calibration. 

For performance evaluation, we used the testing dataset from Ref.~\citenum{beaupre2024quantitative}. It consists of 10 independently-annotated Ca$^{2+}$-imaging videos (of 600 frames each). For each video, an expert had identified the center of the transients by marking the centroid of each mSCT event as a point annotation on the frame corresponding to the maximum relative intensity {\deltaF}. We generated an updated performance evaluation dataset\cite{fan_2026_dataset} in which we manually verified the detections of \astrobeats, \minifinder, and \aqua~on on those 10 Ca$^{2+}$-imaging videos as described in \supref{sec:reverify}. This allowed us to add small and dim transients that were missed by the human annotator in the original testing dataset. The manual verification also allowed to verify transient center alignment with the manual annotations. A catalog of original manual annotations, new machine detections, and manually revised detections is included in this updated performance evaluation dataset Ref.~\citenum{fan_2026_dataset}. This revised dataset is used as the ground truth for comparisons of source-finding performance (see \supref{sec:p_r_f}).

To evaluate segmentation accuracy, we use the original segmentation testset comprising of 69 manually generated masks of mSCT of varying sizes and shapes from Ref.~\citenum{beaupre_2026_18561819}. For each mSCT, two experts had independently annotated masks on $32 \times 32$ pixel crops centered on the frame of maximum relative flux {\deltaF}. 

\subsubsection{Transient Classification}\label{sec:dataset_classes}
To evaluate how source-finding performance may differ across different types of transients, we classify the detections from the revised testing dataset. We provide a catalogue of classification annotations (see Code and Data Availability) for the previously published dataset ~\cite{beaupre_2026_18561819} to better characterize the performance of the algorithms for Ca$^{2+}$ transients that may evolve differently over space or time.\cite{murphy1995mapping,ali2019invivo} We define four broad transients categories in our Ca$^{2+}$-imaging videos: 1) resolved synaptic and 2) dendritic transients, 3) out-of-focus transients and 4) action potentials. 

Synaptic transients are short-lived ($\sim200-300~\mathrm{ms}$), and confined around their point of origin on the dendritic spine ($\sim 1.5~\mu \mathrm{m}$ in maximum length and width). Dendritic transients originate on the dendritic shaft and propagate along it over time, covering a larger area than synaptic transients. These signals are typically associated with Ca$^{2+}$-induced Ca$^{2+}$ release\cite{VERKHRATSKY_1996_CACR}. So-called out-of-focus transients occur outside of the focal plane of the microscope and cannot be properly spatially resolved. Though this type of event cannot be used to study transient properties directly, out-of-focus transients can be used to measure mSCT frequency in a given region. Finally, action potentials are associated with large, widespread fluctuations in the fluorescence signal across all neuronal branches within the field of view (\deltaF $\simeq 100$). Given our specific focus on mSCTs, neurons included in this dataset were treated with Tetrodotoxin (TTX) to suppress action potentials as detailed in Ref.~\citenum{beaupre2024quantitative}. However, in some videos action potentials are still detected, which is flagged by \astrobeats.

\subsection{The \astrobeats~Method} \label{sec:astrobeats}
The \astrobeats~algorithm is designed to generate expert-quality segmentations of mSCTs without the need for manual annotation. Unlike methods such as Refs.~\citenum{yu2024imaging,ting2025imaging}, \astrobeats~does not rely of identifying regions of interest to detect transients, allowing it to detect transients that evolve spatially over time. Drawing inspiration from astronomical transient detection methods, \astrobeats~can be broadly separated into two steps: difference imaging (which includes generating a spatio-temporal mask of dendrites) and source-finding. 

\subsubsection{Difference imaging} \label{sec:astrobeats_background}
We use difference imaging to distinguish mSCTs from foreground dendritic signal by constructing an expected image of the dendrite in the absence of transients. However, both Ca$^{2+}$ diffusion and the Ca$^{2+}$ indicator GCaMP6f dynamics can introduce additional brightness fluctuations along dendrites. \cite{calcnoise,Lohmann2005dendrite,Higley2008dendrite} Therefore, our expected dendrite image must account for these slower, large-scale intensity variations (see \supfigureref{fig:supp_figure_example_fluctuation} for an example of a non-mSCT brightness fluctuation along a dendrite).

To identify dendritic branches, we apply the Rolling Hough Transform (RHT)\cite{rht_clark}, originally developed for detecting astronomical filaments, as an adaptation of the traditional Hough Transform used to detect linear features in images. The RHT generates a map of filament like structures which defines dendritic regions within the Ca$^{2+}$-imaging field of view. Using this map, each frame is decomposed into foreground (dendrite) and background components.

We then construct an expected foreground image for each frame by averaging the dendritic signal across time and locally rescaling it to match the mean intensity of each frame. The expected background is similarly calculated using the average background signal. Both foreground and background estimates are added and smoothed to suppress sharp transitions at the dendrite boundaries, and subsequently subtracted from the raw video on a frame-by-frame basis. This process removes large-scale dendritic brightness variations, isolating local transient activity.

Finally, the residual frames are denoised using a small ($5\times5$ pixels) Gaussian kernel, and the local signal at each pixel ($\sigma$) is expressed in terms of the standard deviation of pixel brightness in its spatiotemporal neighborhood ($5\times5$ pixels $\times ~30$ frames). We refer to the final video after background subtraction and denoising as the transient signal map, $\sigma$. The complete background-subtraction and denoising workflow is illustrated in \autoref{fig:figure_1_intro_schematic}, and mathematical details of the RHT implementation and foreground decomposition are provided in \supref{sec:rht} and \supref{sec:decompose} respectively.

\subsubsection{Source-finding} \label{sec:astrobeats_sources}
Given the diversity of mSCTs event morphologies, which vary in size and expected brightnesses, we use a source-finding procedure that does not rely on any assumption of sphericity or that the sources are point-like, as one might assume in astronomy. We use Density-Based Spatial Clustering of Applications with Noise (DBSCAN)\cite{DBSCAN} to search for groups of bright pixels in our transient signal map. DBSCAN is advantageous in that it allows clusters to be defined based on both cluster size and pixel value, allowing a high diversity of transients to be found.

Clusters are defined in DBSCAN via two main parameters: the maximum separation between neighboring points ($\epsilon$) and the minimum total weight of points ($\mathrm{Min}_\mathrm{pts}$) required to form a cluster. In our implementation, these distances are treated separately in space and time, using $\epsilon_{xy}$ for spatial proximity and $\epsilon_{t}$ for temporal continuity, allowing the algorithm to accommodate transients that evolve over multiple frames. We convert our videos to a list of points defined by the position $(t,x,y)$ of each voxel. We assign weights to each voxel equal to the intensity of that voxel in the transient signal map. To improve computational efficiency, pixels below a brightness threshold, $ \sigma < \mu$, are excluded prior to clustering. We select parameter values according to the optimization methodology described in \supref{sec:optimize}. The parameter values we use in this study are summarized in \suptableref{table:dbscan_params}. 

We separate mSCTs from action potentials by comparing the 2D footprint of each transient to the dendrite location masks obtained through the RHT (see \supref{sec:rht}). Events that occupy a substantial portion of the dendrite ($>50\%$ overlap) are labeled as action potentials and excluded from mSCT detections.

\subsubsection{Segmentation} \label{sec:astrobeats_segmentation}
Detecting transients represents only one aspect of mSCT analysis; characterizing their size, shape, and temporal evolution is equally important, as these morphological features can provide insights into neuronal physiology. As such, segmentation is included as a key task in the \astrobeats~pipeline.

During the difference imaging step (\autoref{sec:astrobeats_background}) we convolve the raw Ca$^{2+}$ image with a Gaussian filter in order to remove sources of Gaussian (white) noise. While this step improves signal strength, it also slightly blurs the boundaries of individual transients. To restore edge definition, we extract a spatio-temporal aperture centered on each detected transient from the raw video, extending 10 frames before and after its temporal peak and covering its spatial footprint. Within this aperture, pixels with raw intensity greater than twice the standard deviation are labeled as part of the transient, creating a binary segmentation mask.

In order to systematically compare transient morphology across the test dataset, including detections absent from the ground truth, we construct a $32\times32$ pixel by 20 time frame aperture around the centroid of each detection. Transient features such as the area, volume or brightness are computed for each aperture (see \suptableref{table:aperture properties}). Detections occurring near image boundaries (within 16 pixels or 10 time frames) are excluded from further analysis to avoid edge artifacts. The resulting set of segmented or manually annotated transient properties provides a basis for quantitative comparison in \autoref{sec:results}. 

\subsection{Intensity-based detection threshold (\minifinder)} \label{sec:flavie_minifinder}
The Intensity-based Detection Threshold (\minifinder) algorithm is an analytical approach used for mSCT detection that can be used as-is or to train a DL model.\cite{beaupre2024quantitative} To address uncorrelated noise bias, \minifinder~subtracts the average intensity of a manually-selected section of the background from each frame of the Ca$^{2+}$-imaging video. The intracellular Ca$^{2+}$ background is determined by manually masking visible dendrites and computing the background within that mask. To detect mSCTs, \minifinder~first compute an estimate of the noise of the baseline fluorescence at each pixel across time. For each pixel, \minifinder~compute the standard deviation (STD) of fluorescence intensity (\deltaF) across time while iteratively removing time points with highest \deltaF. When the estimated STD converges between two consecutive iteration steps, this converged STD is used as the baseline fluorescence. Using this final converged STD, two intensity thresholds are defined: the detection threshold (4 STD) and the segmentation threshold (2 STD). Peaks in intensity above the detection threshold are detected with the MATLAB \texttt{findpeaks} function\cite{matlab_signal}. Any pixel that occurs within a 1500 ms window before or after the occurrence of a peak with a brightness above the segmentation threshold is considered to be part of the same event. To eliminate incorrect detections from instrumental noise or shifts in camera focus, events detected by \minifinder~must be manually verified using a custom graphical user interface (GUI). This re-verification step takes around 30s per transient\cite{beaupre2024quantitative}.

\subsection{Astrocyte Quantitative Analysis (\aqua)} \label{sec:aqua}
The Astrocyte Quantitative Analysis (\aqua)\cite{aqua} algorithm is designed to identify transient events over a large field-of-view, optimized for imaging of spontaneous Ca$^{2+}$ transients in star-shaped astrocyte glial cells. 
The \aqua~algorithm proceeds by first estimating the baseline fluorescence at each pixel. To ensure that the noise follows a Gaussian distribution, variance stabilization is performed, where a square root operation is applied to the data to ensure that the noise variance of the signal does not change with the mean of the background. This results in the background fluorescence following a Gaussian noise distribution; the noise variance of the newly-transformed pixel data then estimated as half of the median of the square of differences between two adjacent values in the time series at the pixel. Active voxels in the image are identified after performing 3D smoothing of the voxels in the image and then identifying groups of spatially-connected outliers (in groups of at least 4).

Bright seed voxels are identified as those with brightness values above a specified threshold, and spatial or temporally adjacent voxels that follow similar temporal evolution of brightness are grouped with these `seed' voxels into super-voxels. The seeds are first extended spatially before extending temporally, and are processed sequentially starting with the brightest seeds. This approach is similar to the \textit{CLEAN} algorithm in radio astronomy where bright sources are removed sequentially from and image until only the estimated background remains\cite{hogbom1974clean}. 

To test the performance of \aqua\ relative to \astrobeats, we vary parameters relating to the initial 3D smoothing of the input video and the detection of spatially connected active super-voxel. We select those that maximize the F$_1$ scores for our test data (described in detail in \supref{sec:p_r_f}). We include the optimization ranges for the parameters together with the optimal value in Supplementary \suptableref{table:aqua}. \aqua~parameters relating to estimating the signal propagation pattern are fixed to their default values. This is consistent with optimization methods used on simulated data\cite{aqua}.

\subsection{Deep Learning-based detection and segmentation}
\label{sec:unet}
Previous work has shown that using the output \minifinder\ to train a supervised neural network can lead to improved detection and segmentation performance of mSCTs\cite{beaupre2024quantitative}. To evaluate this potential, we train a 3D U-Net model\cite{cicek20163dunetlearningdense} with positive unlabeled learning (PU learning) \cite{kilic2012,pu_learning} on the masks generated by both \minifinder~and \astrobeats~and compared their performance for both detection and segmentation tasks (\autoref{sec:results_unet}). PU learning is appropriate when datasets contain labeled positive examples alongside a large number of unlabeled samples, as in the mSCT Ca$^{2+}$-imaging videos.

We train the 3D U-Net using the training set taken from the Ca$^{2+}$-imaging dataset described in \autoref{sec:dataset}\cite{beaupre_2026_18561819}. The training dataset consists of 14 Ca$^{2+}$-imaging videos acquired from different neurons, while the validation dataset consists of 15 Ca$^{2+}$-imaging videos and is kept fixed across all training runs. We use the same training–validation split as in the original publication\cite{beaupre2024quantitative, beaupre_2026_18561819}. Due to the small size of the training data, we measure the average performance by training each model with five random seeds and five subsets of the positive crops from the training set (25 effective folds). The positive $64 \times 64 \times 64$ pixels crops are generated for each event using either \astrobeats~or \minifinder, while unlabeled crops are randomly sampled from the foreground. The 3D U-Net is trained with a ratio of positive labeled to unlabeled items of ($\mathrm{P}-\mathrm{U}$ = 1-64), the ratio that garnered optimal performance in Ref.~\citenum{beaupre2024quantitative}. Detection thresholds of 3D U-Net output are optimized on the segmentation masks of the validation dataset.

\subsection{Performance evaluation}
\label{sec:evaluation}
We evaluate the performance of \astrobeats~, \minifinder\ and \aqua~on the testset (described in\autoref{sec:dataset}) using standard detection and segmentation metrics. For detection, we report Precision, Recall, and the F$_1$ score (see \supref{sec:p_r_f} for detailed equations). Precision measures the ability for the algorithm to reject false events and Recall measures the proportion of confirmed events detected by the algorithm. The F$_1$ score combines both Precision and Recall to offer a more general assessment of detection performance.

For segmentation performance, we use the Diverse Counterfactual Explanations (DiCE)\cite{dice1945measures,dice} similarity score to measures pixel-level overlap between algorithmic and expert-generated masks (see \supref{sec:p_r_f}). Because manual segmentation masks can vary across annotators, we benchmark automated performance against inter-expert agreement rather than perfect DiCE similarity score, following prior work\cite{beaupre2024quantitative}.

\section{Results} \label{sec:results}

\subsection{mSCT detection with \astrobeats} \label{sec:results_manual}
\begin{figure}[htbp!]
\begin{center}
\begin{tabular}{c}
\includegraphics[width=\textwidth, clip]{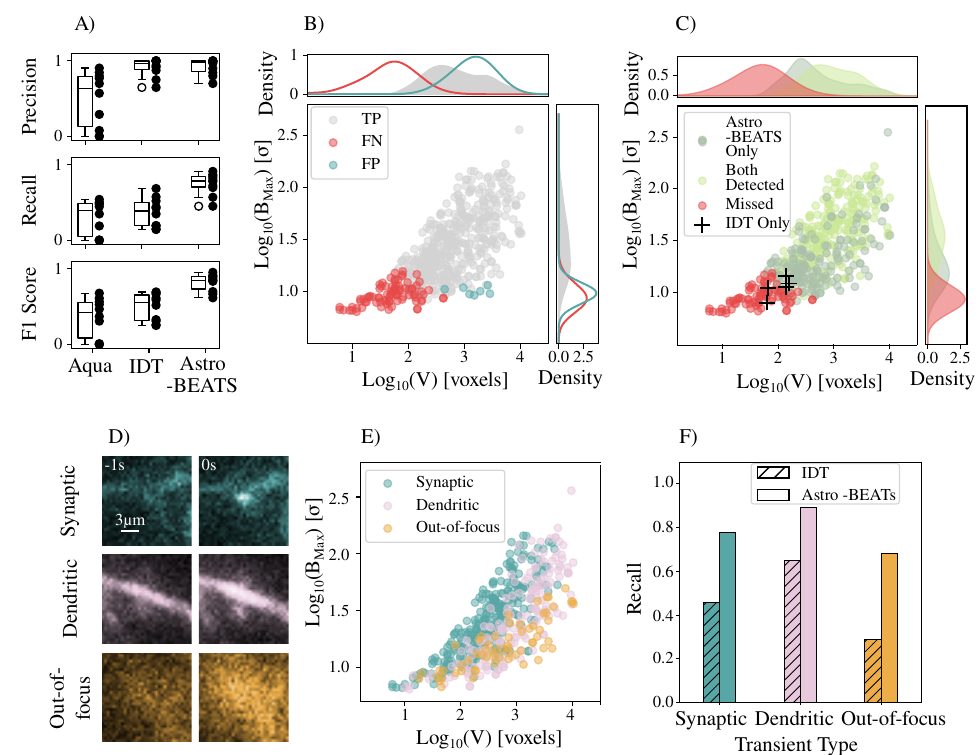}
\end{tabular}
\end{center}
\caption{\label{fig:figure_2_results_detection} A) Precision, Recall and F$_1$ Score for mSCT detection by \astrobeats, \minifinder\, and \aqua\ (N=9 videos). The center line represents the median value while, the boxes represents the 2$^{nd}$ and 3$^{rd}$ quartile range. Whiskers represent the maximum and minimum value and raw data points are plotted beside each box plot. B) Comparison of transients detected by \astrobeats\ compared to manual ground truth annotations (True positive (TP); gray, FN; red, FP; blue). Volume ($V$) is expressed in units of voxels (1 voxel = 0.00256 $\mu$m$^2$ms) and maximum brightness ($B_\mathrm{max}$) is obtained from the transient signal map ($\sigma$). See \suptableref{table:aperture properties} for descriptions of these variables. C) Comparison of transients detected by \astrobeats~and \minifinder\ (detected by both algorithms: green, detected by \astrobeats\ only: grey, detected by \minifinder\ only: black cross, missed by both algorithms: red). D) Example images of out-of-focus, synaptic and dendritic transients during their temporal maximum. We also show the image 1s before the temporal peak of each transient. E) The $V$ and $B_\mathrm{max}$ of each transient types. F) The proportion of manually annotated Ca$^{2+}$ transients recovered by \astrobeats~and \minifinder, for various types of Ca$^{2+}$ transients.}
\end{figure}

We first evaluate the detection performance of mSCTs with \astrobeats\ on the revised test dataset (see \autoref{sec:dataset}). We average the detection performance over 9 videos and find that \astrobeats\ and \minifinder\ exhibit similar Precision scores but \astrobeats\ outperforms the other algorithms for Recall and F1-score  (\autoref{fig:figure_2_results_detection}A). We observe that the smallest and lowest-intensity transients missed by \astrobeats\ (false negatives, FN) are generally also missed by \minifinder\ (\autoref{fig:figure_2_results_detection}B,C). In contrast, slightly larger and brighter transients are frequently detected by \astrobeats\ but remain undetected by \minifinder\ (\autoref{fig:figure_2_results_detection}B). Overall, \astrobeats~detects 74$\%$ of manually identified transients in contrast to around 38$\%$ for \minifinder. Visual inspection of false positive (FP) detections from \astrobeats\ indicates that those are associated in areas characterized by high background noise or overall brightness fluctuation in the video (\autoref{fig:figure_2_results_detection}C). There exists an inevitable trade-off between eliminating noise-generated FP signals and avoiding FN detections: enforcing the detection of dimmer transients may result in incorrectly labeling noise objects as Ca$^{2+}$ transients. 

We next assess the performance of \astrobeats\ for detecting different types of mSCTs. Three classes of transients are identified in the videos: (1) synaptic transients, confined to a small area around the site of calcium entry; (2) dendritic transients, which extend over larger regions along the dendritic shaft; and (3) out-of-focus transients, which lie outside the focal plane and appear as diffuse, blurry regions in the images (\autoref{fig:figure_2_results_detection}D). The testing dataset is manually re-annotated to assign each transient to one of these three classes. These classes differ in both brightness and spatial extent (\autoref{fig:figure_2_results_detection}E). \astrobeats\ outperforms \minifinder\ in detecting all transient types, with particularly strong gains for synaptic and out-of-focus events (\autoref{fig:figure_2_results_detection}F).
The improved performance of \astrobeats\ over the other algorithms can be explained by its capacity to adjust to the fluctuating fluorescence brightness along dendrites and spines through its method of dynamic difference imaging (see \autoref{sec:astrobeats_background}). 

\subsubsection{Segmentation} \label{sec:event_segmentation_compare}
\begin{figure}[htbp!]
\begin{center}
\begin{tabular}{c}
\includegraphics{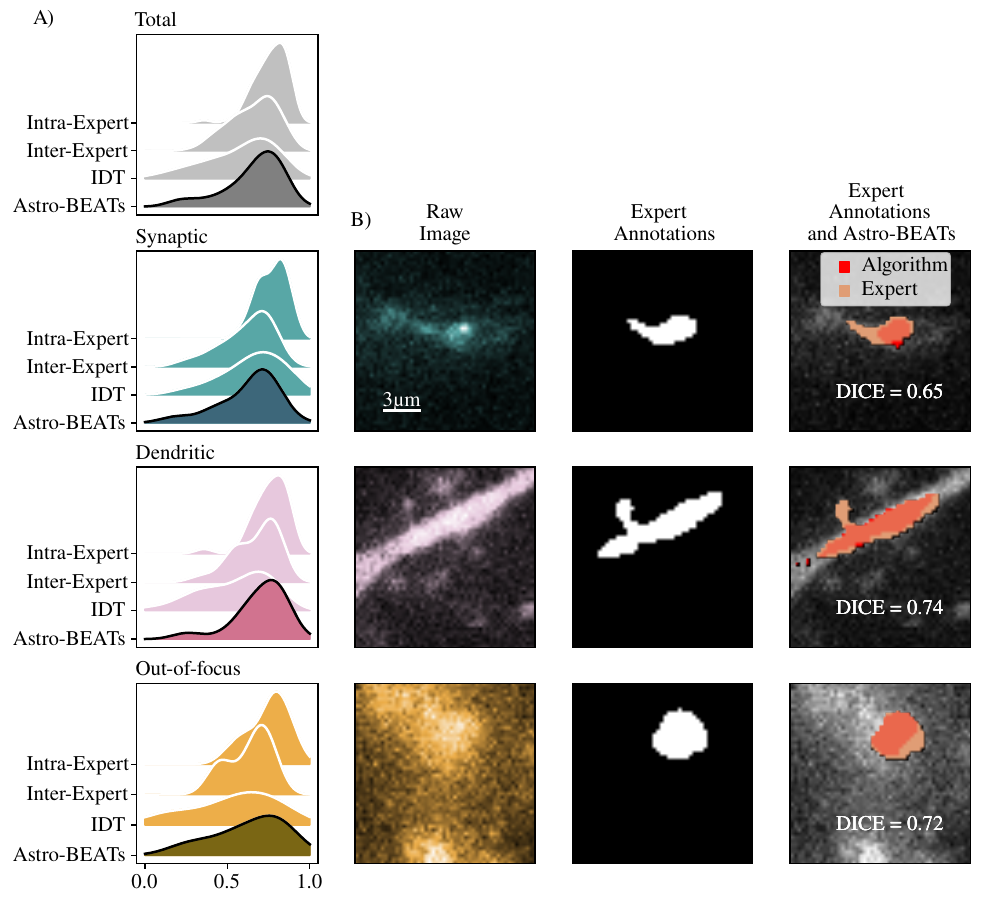}
\end{tabular}
\end{center}
\caption
{\label{fig:figure_3_results_segmentation} 
A) Segmentation DiCE scores calculated over manually segmented transients, using expert segmentation masks as ground truth. Since there is variation between expert annotations, we use inter-expert agreement DiCE scores as a baseline of comparison for segmentation performance (\autoref{sec:event_segmentation_compare}). When comparing segmentation DiCE score across transient types described in \autoref{sec:dataset_classes}, we find that \astrobeats~has comparable DiCE scores to \minifinder~for synaptic transients, while having improved DiCE scores for dendritic and out-of-focus events. C) Example Annotations generated by \astrobeats~compared to expert annotations. We chose example transients with \astrobeats~DiCE segmentation scores closest to the median.}
\end{figure}

In addition to transient detection, we also assess the ability of \astrobeats\ to provide accurate segmentation of the detected mSCTs. We compare the performance of \astrobeats\ with the best performing baseline for the detection task, \minifinder\ as well as inter-expert (median: $\mathrm{DiCE} = 0.69 \pm 0.14$) and intra-expert (median: $\mathrm{DiCE}= 0.77 \pm 0.10$) agreement on manually segmented transients from Ref.~\citenum{beaupre_2026_18561819} (shown in \autoref{fig:figure_3_results_segmentation}A). 

The median DiCE similarity score between \astrobeats~and the expert ground truth annotations is $\mathrm{DiCE}= 0.70 \pm 0.17$ while the previously used \minifinder~obtains a median score of $\mathrm{DiCE}= 0.63 \pm 0.21$. The distribution of DiCE scores for \astrobeats~and \minifinder~are significantly different (the Mann-Whitney test\cite{man1947test}, described in \supref{sec:statistical tests}, returns a value of $p=0.02$ using the null hypothesis that \astrobeats~and \minifinder~DiCE scores are drawn from the same underlying distribution). \astrobeats~attains DiCE scores greater than 0.45 for more transients than \minifinder\ (see \autoref{fig:figure_3_results_segmentation}). Expert annotators state that a DiCE score of $\mathrm{DiCE} > 0.45$ is usually necessary to accurately evaluate the shape of a transient.

As might be expected from its improved performance in detecting different types of transients, we find that \astrobeats~obtains higher DiCE scores compared to \minifinder\ for dendritic transients and out-of-focus events. Notably, \astrobeats~obtains a significantly higher median segmentation DiCE score for dendritic transients  $\mathrm{DiCE}=0.75\pm0.15$, as compared to the median $\mathrm{DiCE}=0.61\pm0.20$ for \minifinder~(Mann-Whitney test, $p=0.002$).The improved segmentation performance for dendritic transients could be beneficial for the analysis of Ca$^{2+}$-induced Ca$^{2+}$ release\cite{VERKHRATSKY_1996_CACR}. \astrobeats~and \minifinder~do not perform significantly differently on synaptic transient segmentations, with both methods obtaining median $\mathrm{DiCE}=0.68\pm0.17$ (Mann-Whitney test $p=0.58$).

\subsubsection{Annotation efficiency}\label{sec:efficiency}
\begin{table}[htbp!]
\caption{Comparison of performance of mSCT segmentation and detection methods. In addition to improvements in annotation and comparable performance in segmentation, \astrobeats~is significantly faster than other approaches.
\label{table:compairalg}}
\begin{center}
\begin{tabular}{ |c|c|c|c|c| }
\hline
\rule[-1ex]{0pt}{3.5ex} Parameter & \astrobeats~& \minifinder~& \aqua~\\ 
\hline\hline
\rule[-1ex]{0pt}{3.5ex} Computation time & $\sim0.6$ s/frame & $\sim30$ s/transient & $\sim 30$ min/video\\
\rule[-1ex]{0pt}{3.5ex} & ($\sim5$ min/video) & ($\sim3$ hr/video) & \\
\hline
\rule[-1ex]{0pt}{3.5ex} Annotation Precision & 0.92 $\pm$ 0.06 & 0.94 $\pm$ 0.10 & 0.56 $\pm$ 0.34\\
\hline
\rule[-1ex]{0pt}{3.5ex} Annotation Recall & 0.75 $\pm$ 0.14 & 0.38 $\pm$ 0.18 & 0.34 $\pm$ 0.22\\
\hline
\rule[-1ex]{0pt}{3.5ex} Annotation F$_1$ score & 0.81 $\pm$ 0.11 & 0.50 $\pm$ 0.17 & 0.39 $\pm$ 0.26\\
\hline
\rule[-1ex]{0pt}{3.5ex} Segmentation DiCE score & 0.66 $\pm$ 0.18 & 0.58 $\pm$ 0.21 & 0.41 $\pm$ 0.24\\
\hline
\rule[-1ex]{0pt}{3.5ex} Required Manual Verification? & No & Yes & No\\
\hline
\end{tabular}

\end{center}
\end{table}
A major advantage of the \astrobeats~method is that it can detect and segment mSCTs in Ca$^{2+}$ without the need for human verification or input. Other approaches that require substantial manual input for training can lead to variations in expert performance over time due to annotation fatigue\cite{vohs2008manual}, especially given that some videos may contain up to $\mathscr{O}(1000)$ events. In contrast, the performance of \astrobeats\ is limited by the speed of its background estimation step (\autoref{sec:astrobeats_background}), after which source identification takes seconds to complete. The \astrobeats\ background estimation pipeline takes $\mathscr{O}(100)$ seconds to complete for each 600-frame video (where each frame has $512 \times 512$ pixels), which represents a factor of ten to a hundred times faster than the other methods discussed in this paper. A summary table of algorithm performance is presented in \autoref{table:compairalg}. 

\subsection{Training Deep Learning Methods on \astrobeats~Annotations}
\label{sec:results_unet}
We next evaluate how \astrobeats\ can be used to train a deep neural network for mSCTs detection and segmentation. As shown previously, DL segmentation algorithms trained on analytical (non-DL) algorithm output can generate improved segmentation masks in comparison to the one used for training\cite{lavoie2020neuronal,beaupre2024quantitative}.  We compare the mSCTs detection and segmentation performance of \astrobeats~to a 3D U-Net (see \autoref{sec:unet}) trained using either \astrobeats~(3D U-Net\textsubscript{AB}) or \minifinder~annotations (3D U-Net\textsubscript{\minifinder}), and \astrobeats~on its own.

\begin{figure}
\begin{center}
\begin{tabular}{c}
\includegraphics[width=0.95\textwidth]{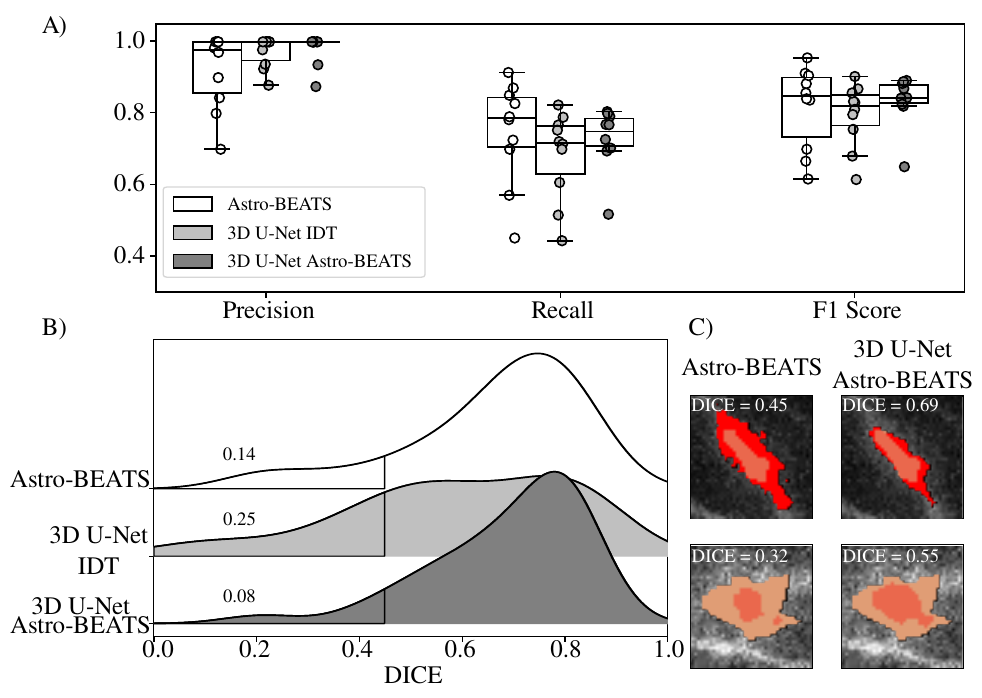}
\end{tabular}
\end{center} 
\caption{\label{fig:figure_4_results_unet}A) Precision, Recall, and F$_1$ score for the optimal performing (see \autoref{sec:unet}) 3D U-Net method trained using \astrobeats~and \minifinder~annotations as ground truth. Despite similar mean values, the variance in both Precision and Recall is reduced for the 3D U-Net\textsubscript{AB} compared to the 3D U-Net\textsubscript{\minifinder} and \astrobeats~only. With the exception of a single video, 3D U-Net\textsubscript{AB} annotations yielded scores of$\mathrm{F}_1>0.8$. B) DiCE score distribution for \astrobeats~, the 3D U-Net\textsubscript{AB}~, and the 3D U-Net\textsubscript{\minifinder}. A cutoff of DiCE $>$ 0.45 is assigned to shown the fraction of segmentations with shapes vastly different from the expert segmentation. Some examples of these non-representative segmentation generated using \astrobeats~and improved using 3D U-Net\textsubscript{AB} are shown in C).}
\end{figure}

The mean values of Precision, Recall and F$_1$ score obtained when using 3D U-Net\textsubscript{AB} do not improve significantly compared to those obtained using \astrobeats~directly ($t$-test score of $p = 0.09$, $p = 0.6$ and $p = 0.5$ for Precision, Recall and F$_1$ respectively, as shown in \autoref{fig:figure_4_results_unet}A. However, the variance in performance is much lower when using 3D U-Net\textsubscript{AB} (\autoref{fig:figure_4_results_unet}). For instance, the Precision obtained using 3D U-Net\textsubscript{AB} has a significantly lower variance of $\sigma = 0.04$ (with a F-test score of $p = 0.006$) than those obtained using \astrobeats~($\sigma = 0.10$). It indicates that the training of the 3D-UNet with \astrobeats\ annotations reduces the likelihood of large variations in the detection quality across individual videos. Although we see lower variance in F$_1$ score and Recall performance for 3D U-Net\textsubscript{AB} compared to \astrobeats~annotations as well, this decrease is not statistically significant(with values of F-test scores of $p = 0.07$ and $p=0.08$ for Recall and F$_1$ respectively) suggesting that improvements in stability are most pronounced for Precision.

Segmentation of mSCTs with the 3D U-Net\textsubscript{AB} results in a significantly higher median DiCE score in comparison to the segmentation with the 3D U-Net\textsubscript{\minifinder} (Mann-Whitney test score of $p=0.01$). Similar to detection metrics, the median segmentation DiCE score obtained using 3D U-Net\textsubscript{AB} do not significantly improve compared to those obtained using \astrobeats~directly. However, for transients that \astrobeats~cannot accurately segment ($\mathrm{DiCE} < 0.45$), 3D U-Net\textsubscript{AB} shows improved performance. As shown in \autoref{fig:figure_4_results_unet}B, only 8$\%$ of the transients in the segmentation testing set scored under $\mathrm{DiCE} = 0.45$ for the 3D U-Net\textsubscript{AB} as opposed to 14$\%$ using \astrobeats~annotations alone. This shows that 3D U-Net\textsubscript{AB} is able to generate segmentation masks indicative of the transient's structure for a higher proportion of transients than the other tested methods. Some examples are shown in \autoref{fig:figure_4_results_unet}C.

\subsection{Generalization} 
\label{sec:out-of-dist}
To test the generalization capability of \astrobeats, we evaluated its performance on an out-of-distribution dataset consisting of Ca$^{2+}$-imaging videos of mSCTs recorded in cultured hippocampal neurons acquired with a different microscope and distinct experimental conditions, (the out-of-distribution dataset described in Ref.~\citenum{beaupre2024quantitative}). We did not re-optimize \astrobeats\ before performing the test. Despite the difference in the dataset used to optimize the original \astrobeats\ and the out-of-distribution dataset shown here, \astrobeats~yielded a high detection accuracy, achieving a value of Precision greater than $\mathrm{P}>0.8$ and value of Recall above $\mathrm{R}>0.6$ (\supfigureref{fig:supp_figure_newset}).

\section{Discussion and Conclusion}
\label{sec:conclusion}
We present \astrobeats, a complete and reliable mSCTs detection and segmentation algorithm developed for the quantitative analysis of Ca$^{2+}$-imaging videos. Since the fluorescence signal of the Ca$^{2+}$ indicator GCaMP6f varies over time, we used the RHT to locate dendrites within our images to estimate their expected brightness and isolate them from foreground transients. The DBSCAN clustering algorithm is used to identify pixels that are associated with mSCTs. We show that our new method can detect a larger proportion of transients in a Ca$^{2+}$-imaging video in comparison to previously used rule-based algorithms, at a significantly faster rate and without incorporating manual verification. This allows \astrobeats~to quickly generate segmentation masks of mSCTs, which can be used as is for quantitative analysis or as ground truth annotations to train a DL model. 
As shown in \autoref{fig:figure_3_results_segmentation}, we find that \astrobeats~is better at segmentation (as indicated by the distribution of DiCE scores) for dendritic and out-of-focus events relative to the other tested models. The improved segmentation masks obtained for all transient types is beneficial for the characterization of transient properties in Ca$^{2+}$-imaging videos. 

A key advantage of \astrobeats~is its strong performance with minimal need for manual re-verification or parameter fine-tuning. As shown in \supref{sec:optimize}, the performance of \astrobeats~is not sensitive to the specific choice of parameters, indicating that the algorithm is robust across a range of input conditions. This property makes \astrobeats~particularly promising for applications to new Ca$^{2+}$-imaging datasets, where imaging parameters, signal-to-noise ratios, or acquisition hardware may differ from the data used to optimize \astrobeats. In \autoref{sec:out-of-dist}, we demonstrate that \astrobeats~can generate detections with high Precision and Recall and segmentation masks that visually resemble the shape of the transient when used out of the box, with no further need for fine-tuning of parameters. This highlights that the algorithm can generalize effectively beyond its initial optimization conditions, supporting its utility as a reliable, unsupervised detection and segmentation framework for diverse Ca$^{2+}$-imaging experiments. 

\section*{Conflicts of Interest}
\label{sec:conflictsofinterest}
The authors declare no competing interests.

\section*{Code and Data Availability}\label{sec:availability}
Open-source python code for \astrobeats~is available at \href{https://github.com/FLClab/Astro-BEATS}{https://github.com/FLClab/Astro-BEATS}. Ca$^{2+}$-imaging data used in this paper is described in detail in Ref.~\citenum{beaupre2024quantitative} and Ref.~\citenum{beaupre_2026_18561819}, and available at \href{https://s3.valeria.science/flclab-calcium/index.html}{https://s3.valeria.science/flclab-calcium/index.html}. A catalogue of Ca$^{2+}$ transient re-annotations and classification introduced in this paper are available in Ref.~\citenum{fan_2026_dataset}.

\section*{Acknowledgments}
\label{sec:acknowledgements}
We thank the Digital Research Alliance of Canada for the compute resources that were used to train and validate the DL models. Funding was provided by grants from the Natural Sciences and Engineering Research Council of Canada (RGPIN-2025-06483 and SMFSU-60768 to R.H., RGPIN-2019-06704 to F.L.C., and RGPIN-2019-06706 to C.G.), the Connaught Fund (R.H.), a Fonds de Recherche Nature et Technologie (FRQNT) Team Grant (2021-PR-284335 to F.L.C and C.G.), Neuronex Initiative (National Science Foundation 2014862, Fond de recherche du Québec - Santé 295824 to F.L.C.), a NSERC-FRQNT NOVA Grant (ALLRP 598002 - 24 to F.L.C., R.H. and C.G) and a New Frontiers in Research Fund Exploration Grant (NFRFE-2020-00933 to R.H., F.L.C., and C.G.). F.L.C. is a Canada Research Chair Tier II (CRC-2019-00126, F.L.C.) and C.G. is a CIFAR AI-Chair. The Dunlap Institute is funded through an endowment established by the David Dunlap family and the University of Toronto. 
The authors at the University of Toronto acknowledge that the land on which the University of Toronto is built is the traditional territory of the Haudenosaunee, and most recently, the territory of the Mississaugas of the New Credit First Nation. They are grateful to have the opportunity to work in the community, on this territory. A.B. was supported by scholarships from NSERC, F.B. was supported by scholarships from FRQNT, NeuroQuébec, and the Québec Bio-Imaging Network. A.B., F.B. and T.W. were awarded excellence scholarships from the FRQNT strategic cluster UNIQUE.


\bibliography{report}   

@article{Higley2008dendrite,
title = {Calcium Signaling in Dendrites and Spines: Practical and Functional Considerations},
journal = {Neuron},
volume = {59},
number = {6},
pages = {902-913},
year = {2008},
issn = {0896-6273},
doi = {https://doi.org/10.1016/j.neuron.2008.08.020},
url = {https://www.sciencedirect.com/science/article/pii/S0896627308007447},
author = {Michael J. Higley and Bernardo L. Sabatini}
}

@article{ali2019invivo,
author = {Farhan Ali and Alex C. Kwan},
title = {{Interpreting in vivo calcium signals from neuronal cell bodies, axons, and dendrites: a review}},
volume = {7},
journal = {Neurophotonics},
number = {1},
publisher = {SPIE},
pages = {011402},
year = {2019},
doi = {10.1117/1.NPh.7.1.011402},
URL = {https://doi.org/10.1117/1.NPh.7.1.011402}
}

@article{Lohmann2005dendrite,
title = {Regulation of dendritic growth and plasticity by local and global calcium dynamics},
journal = {Cell Calcium},
volume = {37},
number = {5},
pages = {403-409},
year = {2005},
note = {Calcium in the function of the nervous system: New implications},
issn = {0143-4160},
doi = {https://doi.org/10.1016/j.ceca.2005.01.008},
url = {https://www.sciencedirect.com/science/article/pii/S0143416005000242},
author = {Christian Lohmann and Rachel O.L. Wong}
}

@article{yu2024imaging,
author = {Yiyi Yu and Liam M. Adsit and Ikuko  T. Smith},
title = {{Comprehensive software suite for functional analysis and synaptic input mapping of dendritic spines imaged in vivo}},
volume = {11},
journal = {Neurophotonics},
number = {2},
publisher = {SPIE},
pages = {024307},
year = {2024},
doi = {10.1117/1.NPh.11.2.024307},
URL = {https://doi.org/10.1117/1.NPh.11.2.024307}
}

@article{ting2025imaging,
author = {Li, Ting and Wei, Hengqi and Zhao, Cong Jian},
year = {2025},
month = {05},
pages = {01011},
title = {Synapse Calcium Image Analysis (SCIA): A Groundbreaking Toolkit for Real-Time Analysis of synaptic Calcium Signals},
volume = {174},
journal = {BIO Web of Conferences},
doi = {10.1051/bioconf/202517401011}
}

@article{murphy1995mapping,
  title={Mapping miniature synaptic currents to single synapses using calcium imaging reveals heterogeneity in postsynaptic output},
  author={Murphy, Timothy H and Baraban, Jay M and Wier, W Gil},
  journal={Neuron},
  volume={15},
  number={1},
  pages={159--168},
  year={1995},
  publisher={Cell Press}
}

@article{kuhn1955hungarian,
  title={The Hungarian method for the assignment problem},
  author={Kuhn, Harold W},
  journal={Naval research logistics quarterly},
  volume={2},
  number={1-2},
  pages={83--97},
  year={1955},
  publisher={Wiley Online Library}
}

@article{agarwal2017transient,
  title={Transient opening of the mitochondrial permeability transition pore induces microdomain calcium transients in astrocyte processes},
  author={Agarwal, Amit and Wu, Pei-Hsun and Hughes, Ethan G and Fukaya, Masahiro and Tischfield, Max A and Langseth, Abraham J and Wirtz, Denis and Bergles, Dwight E},
  journal={Neuron},
  volume={93},
  number={3},
  pages={587--605},
  year={2017},
  publisher={Elsevier}
}

@article{srinivasan2015ca2,
  title={Ca2+ signaling in astrocytes from Ip3r2-/- mice in brain slices and during startle responses in vivo},
  author={Srinivasan, Rahul and Huang, Ben S and Venugopal, Sharmila and Johnston, April D and Chai, Hua and Zeng, Hongkui and Golshani, Peyman and Khakh, Baljit S},
  journal={Nature neuroscience},
  volume={18},
  number={5},
  pages={708--717},
  year={2015},
  publisher={Nature Publishing Group US New York}
}

@article{cuny2022microscopy,
    author = {Cuny, Andreas P. and Schlottmann, Fabian P. and Ewald, Jennifer C. and Pelet, Serge and Schmoller, Kurt M.},
    title = {Live cell microscopy: From image to insight},
    journal = {Biophysics Reviews},
    volume = {3},
    number = {2},
    pages = {021302},
    year = {2022},
    month = {04},
    issn = {2688-4089},
    doi = {10.1063/5.0082799},
    url = {https://doi.org/10.1063/5.0082799},
    eprint = {https://pubs.aip.org/aip/bpr/article-pdf/doi/10.1063/5.0082799/19804528/021302_1_online.pdf},
}

@article{hogbom1974clean,
    author = "Hogbom, J. A.",
    title = "{Aperture Synthesis with a Non-Regular Distribution of Interferometer Baselines}",
    journal = "Astron. Astrophys. Suppl. Ser.",
    volume = "15",
    pages = "417--426",
    year = "1974"
}

@misc{fan_2026_dataset,
  author       = {Fan, Bolin and
                  Bilodeau, Anthony and
                  Beaupré, Frédéric and
                  Lavoie-Cardinal, Flavie and
                  Hlozek, Renee},
  title        = {Zenodo data link for: Ca2+ transient detection and segmentation with the Astronomically motivated algorithm for Background Estimation And Transient Segmentation (Astro-BEATS): \href{https://doi.org/10.5281/zenodo.18601510}{10.5281/zenodo.18601510}},
  month        = feb,
  year         = 2026,
  publisher    = {Zenodo},
  doi          = {10.5281/zenodo.18601510},
  url          = {https://doi.org/10.5281/zenodo.18601510},
}

@article{dotti2024deep,
  title={A Deep Learning-Based Approach for Efficient Detection and Classification of Local Ca$^{2+}$ Release Events in Full-Frame Confocal Imaging},
  author={Dotti, Prisca and Fernandez-Tenorio, Miguel and Janicek, Radoslav and M{\'a}rquez-Neila, Pablo and Wullschleger, Marcel and Sznitman, Raphael and Egger, Marcel},
  journal={Cell Calcium},
  pages={102893},
  year={2024},
  publisher={Elsevier}
}

@misc{beaupre_2026_18561819,
  author       = {Beaupré, Frédéric and
                  Bilodeau, Anthony and
                  Wiesner, Theresa and
                  Leclerc, Gabriel and
                  Gagné, Christian and
                  De Koninck, Paul and
                  Lavoie-Cardinal, Flavie},
  title        = {Quantitative Analysis of Miniature Synaptic
                   Calcium Transients using Positive Unlabeled Deep
                   Learning
                  },
  month        = feb,
  year         = 2026,
  publisher    = {Zenodo},
  doi          = {10.5281/zenodo.18561819},
  url          = {https://doi.org/10.5281/zenodo.18561819},
}

@inproceedings{xu2019u,
  title={U-net with optimal thresholding for small blob detection in medical images},
  author={Xu, Yanzhe and Gao, Fei and Wu, Teresa and Bennett, Kevin M and Charlton, Jennifer R and Sarkar, Suryadipto},
  booktitle={2019 IEEE 15th International Conference on Automation Science and Engineering (CASE)},
  pages={1761--1767},
  year={2019},
  organization={IEEE}
}

@article{aqua,
author = {Wang, Yizhi and DelRosso, Nicole and Vaidyanathan, Trisha and Cahill, Michelle and Reitman, Michael and Pittolo, Silvia and Mi, Xuelong and Yu, Guoqiang and Poskanzer, Kira},
year = {2019},
month = {11},
pages = {1-9},
title = {Accurate quantification of astrocyte and neurotransmitter fluorescence dynamics for single-cell and population-level physiology},
volume = {22},
journal = {Nature Neuroscience},
doi = {10.1038/s41593-019-0492-2}
}

@Article{kilic2012,
author={K{\i}l{\i}{\c{c}}, Cumhur
and Tan, Mehmet},
title={Positive unlabeled learning for deriving protein interaction networks},
journal={Network Modeling Analysis in Health Informatics and Bioinformatics},
year={2012},
month={Sep},
day={01},
volume={1},
number={3},
pages={87-102},
issn={2192-6670},
doi={10.1007/s13721-012-0012-8},
url={https://doi.org/10.1007/s13721-012-0012-8}
}

@article{pu_learning,
  title={Learning from positive and unlabeled data: A survey},
  author={Bekker, Jessa and Davis, Jesse},
  journal={Machine Learning},
  volume={109},
  pages={719--760},
  year={2020},
  publisher={Springer}
}

@inproceedings{dice,
author = {Mothilal, Ramaravind K. and Sharma, Amit and Tan, Chenhao},
title = {Explaining machine learning classifiers through diverse counterfactual explanations},
year = {2020},
isbn = {9781450369367},
publisher = {Association for Computing Machinery},
address = {New York, NY, USA},
url = {https://doi.org/10.1145/3351095.3372850},
doi = {10.1145/3351095.3372850},
booktitle = {Proceedings of the 2020 Conference on Fairness, Accountability, and Transparency},
pages = {607–617},
numpages = {11},
location = {Barcelona, Spain},
series = {FAT* '20}
}

@article{VERKHRATSKY_1996_CACR,
title = {Calcium-induced calcium release in neurones},
journal = {Cell Calcium},
volume = {19},
number = {1},
pages = {1-14},
year = {1996},
issn = {0143-4160},
doi = {https://doi.org/10.1016/S0143-4160(96)90009-3},
url = {https://www.sciencedirect.com/science/article/pii/S0143416096900093},
author = {A. Verkhratsky and A. Shmigol}
}

@article{murthy2000dynamics,
  title={Dynamics of dendritic calcium transients evoked by quantal release at excitatory hippocampal synapses},
  author={Murthy, Venkatesh N and Sejnowski, Terrence J and Stevens, Charles F},
  journal={Proceedings of the National Academy of Sciences},
  volume={97},
  number={2},
  pages={901--906},
  year={2000},
  publisher={National Acad Sciences}
}

@article{masias2012review,
  title={A review of source detection approaches in astronomical images},
  author={Masias, M and Freixenet, J and Llad{\'o}, X and Peracaula, M},
  journal={Monthly Notices of the Royal Astronomical Society},
  volume={422},
  number={2},
  pages={1674--1689},
  year={2012},
  publisher={Blackwell Publishing Ltd Oxford, UK}
}

@article{zackay2016proper,
  title={Proper image subtraction—optimal transient detection, photometry, and hypothesis testing},
  author={Zackay, Barak and Ofek, Eran O and Gal-Yam, Avishay},
  journal={The Astrophysical Journal},
  volume={830},
  number={1},
  pages={27},
  year={2016},
  publisher={IOP Publishing}
}

@ARTICLE{blobcat2012,
       author = {{Hales}, C.~A. and {Murphy}, T. and {Curran}, J.~R. and {Middelberg}, E. and {Gaensler}, B.~M. and {Norris}, R.~P.},
        title = "{BLOBCAT: software to catalogue flood-filled blobs in radio images of total intensity and linear polarization}",
      journal = {Monthly Notices of the Royal Astronomical Society},
         year = 2012,
        month = sep,
       volume = {425},
       number = {2},
        pages = {979-996},
          doi = {10.1111/j.1365-2966.2012.21373.x},
archivePrefix = {arXiv},
       eprint = {1205.5313},
 primaryClass = {astro-ph.IM},
       adsurl = {https://ui.adsabs.harvard.edu/abs/2012MNRAS.425..979H}
}

@ARTICLE{hancock2012,
       author = {{Hancock}, P.~J. and {Murphy}, T. and {Gaensler}, B.~M. and {Hopkins}, A. and {Curran}, J.~R.},
        title = "{Compact continuum source finding for next generation radio surveys}",
      journal = {Monthly Notices of the Royal Astronomical Society},
         year = 2012,
        month = may,
       volume = {422},
       number = {2},
        pages = {1812-1824},
          doi = {10.1111/j.1365-2966.2012.20768.x},
archivePrefix = {arXiv},
       eprint = {1202.4500},
 primaryClass = {astro-ph.IM},
       adsurl = {https://ui.adsabs.harvard.edu/abs/2012MNRAS.422.1812H}
}

@misc{matlab_signal,
      title={GSPBOX: A toolbox for signal processing on graphs}, 
      author={Nathanaël Perraudin and Johan Paratte and David Shuman and Lionel Martin and Vassilis Kalofolias and Pierre Vandergheynst and David K. Hammond},
      year={2016},
      eprint={1408.5781},
      archivePrefix={arXiv},
      primaryClass={cs.IT},
      url={https://arxiv.org/abs/1408.5781}, 
}

@article{td_2,
title={The Completeness and Reliability of Threshold and False-discovery Rate Source Extraction Algorithms for Compact Continuum Sources}, volume={29}, DOI={10.1071/AS11026}, number={3}, journal={Publications of the Astronomical Society of Australia}, publisher={Cambridge University Press}, author={Huynh, M. T. and Hopkins, A. and Norris, R. and Hancock, P. and Murphy, T. and Jurek, R. and Whiting, M.}, year={2012}, pages={229–243}}

@article{big_data,
author = {Feigelson, Eric D. and Babu, G. Jogesh},
title = {Big data in astronomy},
journal = {Significance},
volume = {9},
number = {4},
pages = {22-25},
doi = {https://doi.org/10.1111/j.1740-9713.2012.00587.x},
url = {https://rss.onlinelibrary.wiley.com/doi/abs/10.1111/j.1740-9713.2012.00587.x},
eprint = {https://rss.onlinelibrary.wiley.com/doi/pdf/10.1111/j.1740-9713.2012.00587.x},
year = {2012}
}

@article{td_1,
    author = {Bramich, D. M.},
    title = "{A new algorithm for difference image analysis}",
    journal = {Monthly Notices of the Royal Astronomical Society: Letters},
    volume = {386},
    number = {1},
    pages = {L77-L81},
    year = {2008},
    month = {05},
    issn = {1745-3925},
    doi = {10.1111/j.1745-3933.2008.00464.x},
    url = {https://doi.org/10.1111/j.1745-3933.2008.00464.x},
    eprint = {https://academic.oup.com/mnrasl/article-pdf/386/1/L77/3909938/386-1-L77.pdf},
}

@ARTICLE{DBSCAN,
       author = {{Sander}, J{\"o}rg and {Ester}, Martin and {Kriegel}, Hans-Peter and {Xu}, Xiaowei},
        title = "{Density-Based Clustering in Spatial Databases: The Algorithm GDBSCAN and Its Applications}",
      journal = {Data Mining and Knowledge Discovery},
         year = 1998,
        month = jun,
       volume = {2},
       number = {2},
        pages = {169},
          doi = {10.1023/A:1009745219419},
       adsurl = {https://ui.adsabs.harvard.edu/abs/1998DMKD....2..169S}
}

@article{vohs2008manual,
author = {Vohs, Kathleen and Baumeister, Roy and Schmeichel, Brandon and Twenge, Jean and Nelson, Noelle and Tice, Dianne},
year = {2008},
month = {06},
pages = {883-98},
title = {Making Choices Impairs Subsequent Self-Control: A Limited-Resource Account of Decision Making, Self-Regulation, and Active Initiative},
volume = {94},
journal = {Journal of personality and social psychology},
doi = {10.1037/0022-3514.94.5.883}
}

@article{man1947test,
 ISSN = {00034851},
 URL = {http://www.jstor.org/stable/2236101},
 author = {H. B. Mann and D. R. Whitney},
 journal = {The Annals of Mathematical Statistics},
 number = {1},
 pages = {50--60},
 publisher = {Institute of Mathematical Statistics},
 title = {On a Test of Whether one of Two Random Variables is Stochastically Larger than the Other},
 urldate = {2026-01-12},
 volume = {18},
 year = {1947}
}

@article {shapiro/wilk,
    AUTHOR = {Shapiro, S. S. and Wilk, M. B.},
     TITLE = {An analysis of variance test for normality: {C}omplete
              samples},
   JOURNAL = {Biometrika},
  FJOURNAL = {Biometrika},
    VOLUME = {52},
      YEAR = {1965},
     PAGES = {591--611},
      ISSN = {0006-3444,1464-3510},
   MRCLASS = {62.25 (62.71)},
  MRNUMBER = {205384},
MRREVIEWER = {N.\ L.\ Johnson},
       DOI = {10.1093/biomet/52.3-4.591},
       URL = {https://doi.org/10.1093/biomet/52.3-4.591},
}

@article{calcnoise,
title = {Noise analysis of cytosolic calcium image data},
journal = {Cell Calcium},
volume = {86},
pages = {102152},
year = {2020},
issn = {0143-4160},
doi = {https://doi.org/10.1016/j.ceca.2019.102152},
url = {https://www.sciencedirect.com/science/article/pii/S0143416019302210},
author = {Divya Swaminathan and George D. Dickinson and Angelo Demuro and Ian Parker}
}

@article{gcamp,
author = {Chen, Tsai-Wen and Wardill, Trevor and Sun, Yi and Pulver, Stefan and Renninger, Sabine and Baohan, Amy and Schreiter, Eric and Kerr, Rex and Orger, Michael and Jayaraman, Vivek and Looger, Loren and Svoboda, Karel and Kim, Douglas},
year = {2013},
month = {07},
pages = {295-300},
title = {Ultrasensitive fluorescent proteins for imaging neuronal activity},
volume = {499},
journal = {Nature},
doi = {10.1038/nature12354}
}

@article {beaupre2024quantitative,
author ="Beaupré, Frédéric and Bilodeau, Anthony and Wiesner, Theresa and Leclerc, Gabriel and Lemieux, Mado and Nadeau, Gabriel and Castonguay, Katrine and Fan, Bolin and Labrecque, Simon and Hložek, Renée and De Koninck, Paul and Gagné, Christian and Lavoie-Cardinal, Flavie",
title  ="Quantitative analysis of miniature synaptic calcium transients using positive unlabeled deep learning",
journal  ="Digital Discovery",
year  ="2025",
volume  ="4",
issue  ="1",
pages  ="105-119",
publisher  ="RSC",
doi  ="10.1039/D4DD00197D",
url  ="http://dx.doi.org/10.1039/D4DD00197D"}

@misc{cicek20163dunetlearningdense,
      title={3D U-Net: Learning Dense Volumetric Segmentation from Sparse Annotation}, 
      author={Ozgun Cicek and Ahmed Abdulkadir and Soeren S. Lienkamp and Thomas Brox and Olaf Ronneberger},
      year={2016},
      eprint={1606.06650},
      archivePrefix={arXiv},
      primaryClass={cs.CV},
      url={https://arxiv.org/abs/1606.06650}, 
}

@MISC{rht_clark,
       author = {{Clark}, Susan E. and {Peek}, Josh and {Putman}, Mary and {Schudel}, Lowell and {Jaspers}, Rutger},
        title = "{RHT: Rolling Hough Transform}",
 howpublished = {Astrophysics Source Code Library, record ascl:2003.005},
         year = 2020,
        month = mar,
          eid = {ascl:2003.005},
        pages = {ascl:2003.005},
archivePrefix = {ascl},
       eprint = {2003.005},
       adsurl = {https://ui.adsabs.harvard.edu/abs/2020ascl.soft03005C}
}

@article{andreae2015spontaneous,
  title={Spontaneous neurotransmitter release shapes dendritic arbors via long-range activation of NMDA receptors},
  author={Andreae, Laura C and Burrone, Juan},
  journal={Cell reports},
  volume={10},
  number={6},
  pages={873--882},
  year={2015},
  publisher={Elsevier}
}

@article{kerr2008imaging,
  title={Imaging in vivo: watching the brain in action},
  author={Kerr, Jason ND and Denk, Winfried},
  journal={Nature Reviews Neuroscience},
  volume={9},
  number={3},
  pages={195--205},
  year={2008},
  publisher={Nature Publishing Group UK London}
}

@article{grienberger2012imaging,
  title={Imaging calcium in neurons},
  author={Grienberger, Christine and Konnerth, Arthur},
  journal={Neuron},
  volume={73},
  number={5},
  pages={862--885},
  year={2012},
  publisher={Elsevier}
}

@article{kavalali2015mechanisms,
  title={The mechanisms and functions of spontaneous neurotransmitter release},
  author={Kavalali, Ege T},
  journal={Nature Reviews Neuroscience},
  volume={16},
  number={1},
  pages={5--16},
  year={2015},
  publisher={Nature Publishing Group UK London}
}

@article{reese2016single,
  title={Single synapse evaluation of the postsynaptic NMDA receptors targeted by evoked and spontaneous neurotransmission},
  author={Reese, Austin L and Kavalali, Ege T},
  journal={Elife},
  volume={5},
  pages={e21170},
  year={2016},
  publisher={eLife Sciences Publications, Ltd}
}

@article{dice1945measures,
  title={Measures of the amount of ecologic association between species},
  author={Dice, Lee R},
  journal={Ecology},
  volume={26},
  number={3},
  pages={297--302},
  year={1945},
  publisher={JSTOR}
}

@article{lavoie2020neuronal,
  title={Neuronal activity remodels the F-actin based submembrane lattice in dendrites but not axons of hippocampal neurons},
  author={Lavoie-Cardinal, Flavie and Bilodeau, Anthony and Lemieux, Mado and Gardner, Marc-Andr{\'e} and Wiesner, Theresa and Laram{\'e}e, Gabrielle and Gagn{\'e}, Christian and De Koninck, Paul},
  journal={Scientific reports},
  volume={10},
  number={1},
  pages={11960},
  year={2020},
  publisher={Nature Publishing Group UK London}
}
\bibliographystyle{spiejour}   

\section*{Author Contributions}
B.F. and R.H. designed the \astrobeats\ analysis pipeline. B.F. implemented the \astrobeats\ algorithm, the \aqua and \minifinder\ baselines, performed the analysis, validated the results, and generated the updated detection and classification annotations. T.W. and F.L.C. validated the performance of \astrobeats\ and of the updated annotations. F.L.C. designed the user-study. A.B. and F.B. implemented the D.L. pipelines. F.L.C, C.G., A.B. and F.B. validated the DL pipelines. A.B., F.B and B.F. analyzed the results of the DL models. B.F., R.H., F.L.C. and A.B. wrote the manuscript. 
\clearpage

\cleardoublepage
\renewcommand\appendixpagename{Supplementary Material}
\begin{appendices}
\label{Supplimentary}
\renewcommand{\appendixname}{}
\appendixtitleon    
\appendixtitletocon 
\titleformat{\section}{\bfseries}{\thesection.}{0pt}{~}{} 

\renewcommand{\figurename}{Supplementary Fig}
\renewcommand{\tablename}{Supplementary Table}
\newcounter{suppfig}
\setcounter{suppfig}{1}
\newcounter{supptab}
\setcounter{supptab}{1}

\renewcommand\thefigure{\arabic{suppfig}}
\renewcommand\thetable{\arabic{supptab}}
\setcounter{page}{1}

\section{Rolling Hough Transform} \label{sec:rht}
Every line in two dimensional (2D) space may be uniquely parameterized according to: 

\begin{equation}\rho = x\cos(\theta) + y\sin(\theta),\end{equation}

where $\theta$ is the angle from normal and $\rho$ is the minimum Euclidean distance between the line and the origin in Cartesian space. The HT converts an image from Cartesian space ($x$,$y$) to Hough space ($\rho$,$\theta$) by counting the number of sufficiently bright pixels along each line parameterized by $\rho$ and $\theta$ through an image. Bright lines within an image will appear as peaks in Hough space.  The RHT method can detect straight line segments as well as curved filaments. In our application, the image is first convolved with a tophat filter (diameter = 41 pixels in our case) to remove sharp features. This smoothed image is then subtracted from the original image to remove large-scale structures. On this new image containing only sharp edges, the RHT selects a circular region of radius ($r = 6$) centered on a pixel $x,y$. A HT is performed for $\rho = 0$, centered on the central pixel of this region. This HT returns a score, $R(\theta,x,y)$, for the degree of linear structure at each $\theta$ through that pixel. The process is repeated for every pixel $(x,y)$ in a image, generating a map of linear structure, as shown in the top row of \autoref{fig:figure_1_intro_schematic}.

\section{Decomposition} \label{sec:decompose}
We apply the RHT to the median intensity time-projection of all frames with a maximum intensity below $5\sigma$ of the mean video intensity. This generates a 2D map of dendrite locations, $R(x,y)$, within the Ca$^{2+}$-imaging video. According to this map, each Ca$^{2+}$-imaging video may be split into foreground ($F_{dendrite}$) and background ($F_{background}$) components: \begin{eqnarray}
    F(t,x,y) &=& F_{dendrite}(t,x,y) + F_{background} (t,x,y)\nonumber \\
&\mathrm{where}&F_{dendrite} (t,x,y)\,\,\,\,\,=
    \left\{ \begin{array}{ll}
        F(t,x,y) & \mathrm{if}\,\,R(x,y) = 1 \\
        0 & \mathrm{if}\,\, R(x,y) = 0\\
    \end{array} \right. \nonumber \\
    &\mathrm{and}&
    F_{background} (t,x,y)=
    \left\{ \begin{array}{ll}
        0 & \mathrm{if}\,\,  R(x,y) = 1 \\
        F(t,x,y) & \mathrm{if}\,\,  R(x,y) = 0
    \end{array} \right. \nonumber\\
\label{eq:background}
\end{eqnarray} After decomposing the Ca$^{2+}$ video into foreground and background components, a generic foreground image ($\langle F_{dendrite}\rangle(x,y)$) is constructed by first taking the mean intensity of the foreground ($F_{dendrite}(t,x,y)$ in \autoref{eq:background}) across all frames at each pixel. For every given frame, the mean intensity of $\langle F_{dendrite}\rangle(x,y)$ is rescaled to be equal to the mean foreground intensity of the target frame, and the 2 frames immediately before or after it: $[t-1,t+1]$. The generic background image ($\langle F_{background}\rangle(x,y)$) is similarly obtained by taking the mean intensity of the background ($F_{background}(t,x,y)$). $\langle F_{background}\rangle(x,y)$ is scaled to match the median background intensity of frames $[t-1,t+1]$, rather than the mean. This suppresses foreground fluctuations that may affect nearby background pixels. Both adjusted images are then smoothed with a $10\times10$-pixel by 1-frame tophat filter and subsequently subtracted from the raw video ($F(t,x,y)$) on a frame-by-frame basis.

\section{Optimizing \astrobeats~performance} \label{sec:optimize} 
We optimize the \astrobeats~parameters as follows: for a subset of the DBSCAN parameters with the range defined in \suptableref{table:dbscan_params}. We optimize the parameters on one video in a set of 10 test videos. We find the parameter combination that yields the greatest F$_1$ score on this video, as defined in \autoref{eq:F_beta}). These parameters are next used to assess the performance of \astrobeats~ on the other 9 videos.

\begin{figure}[htbp!]
\begin{center}
\includegraphics{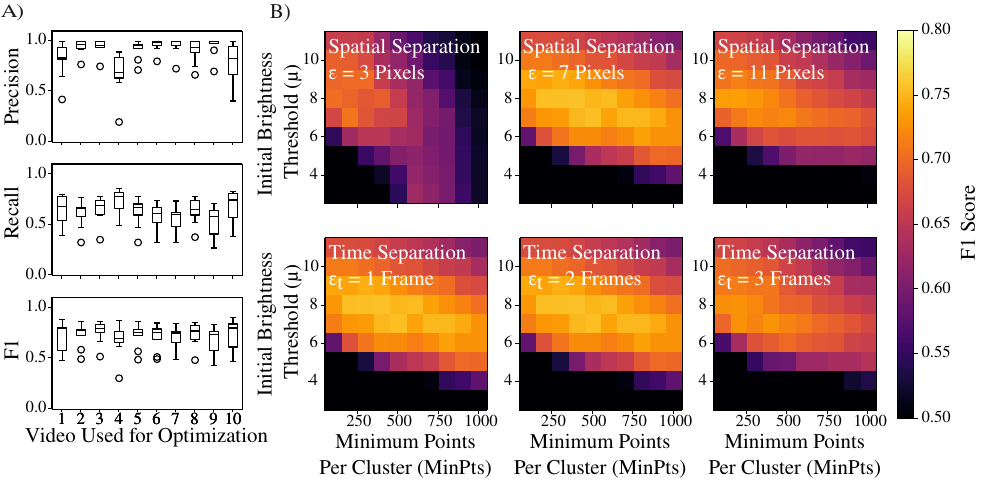}
\end{center}
\caption
{\label{fig:supp_figure_optimization} A) Optimization of the DBSCAN parameters. The range of DBSCAN parameters that yield good F$_1$ scores are shown in \autoref{table:dbscan_params}. We find that for a range of $400< \mathrm{Min}_\mathrm{pts} < 700$ and $5\sigma < \mu < 9\sigma$, we obtain F$_1 > 0.7$, which we consider to be acceptable. B) Precision, Recall and F$_1$ scores obtained by finding the DBSCAN parameters that achieve the highest F$_1$ score for one video, and applied over the remaining 9 test videos.}
\addtocounter{suppfig}{1}
\end{figure}

We see that the F$_1$ score remains above 0.7 for a range of parameters, indicating that \astrobeats~performance is not strongly dependent on parameter choice (\supfigureref{fig:supp_figure_optimization}). \astrobeats~yields an average F$_1$ score of around $F_1=0.78 \pm 0.07$ across the 10 videos used for optimization (\supfigureref{fig:supp_figure_optimization}A). This shows that \astrobeats~performance is robust when it is optimized on a subset of observations. The \astrobeats~parameters used in this paper were optimized using Video 5. This is because the F$_1$ score performance obtained using these parameters is closest to the median F$_1$ score across all candidate optimization videos, reducing the risk of over or underfitting.

\section{Reverification} \label{sec:reverify}
\begin{figure}[htbp!]
\begin{center}
\includegraphics{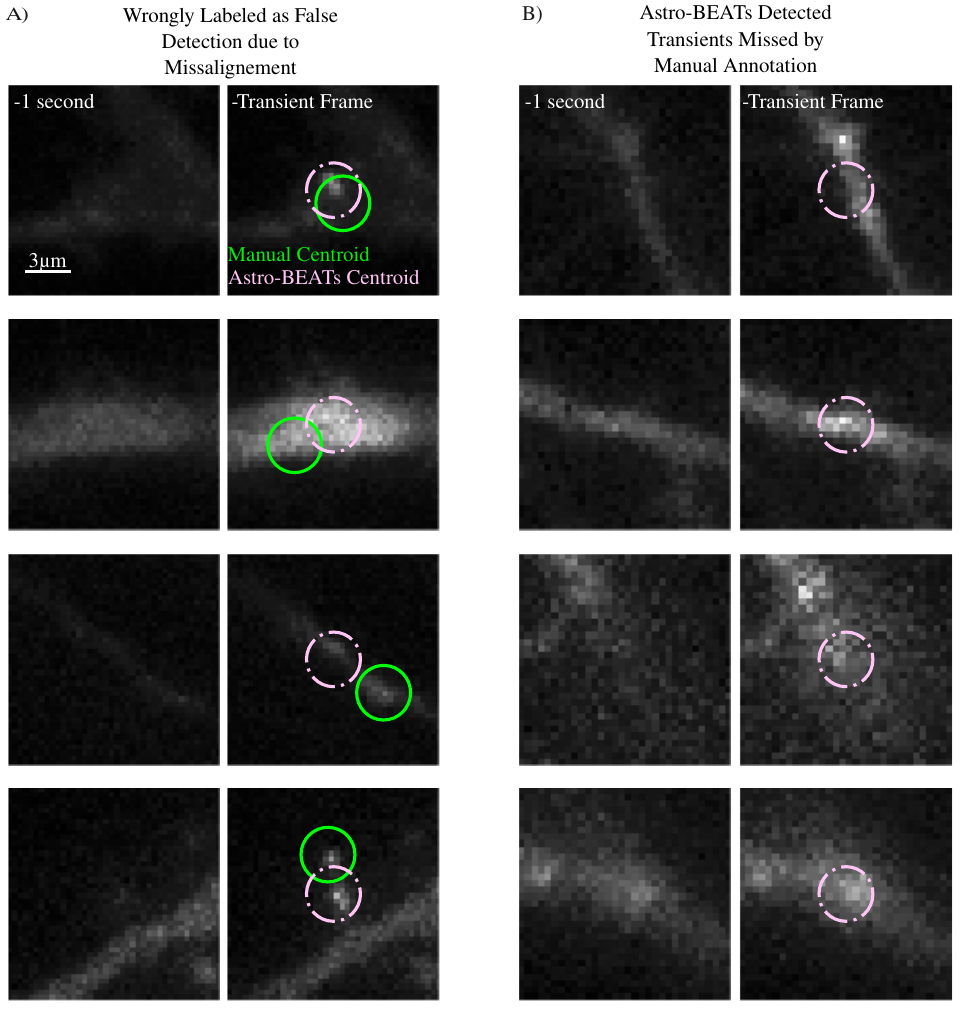}
\end{center}
\caption
{\label{fig:supp_figure_reverify_examples} A) TP detections made by \astrobeats~that were classified as FP or FN due to misalignment error. B) Small and dim transients detected using \astrobeats~but missed through manual annotation of the original testing dataset.}
\addtocounter{suppfig}{1}
\end{figure}

We manually inspected the FP and FN detections from \aqua, \minifinder, and \astrobeats~to identify two types of miss-match between expert and machine annotations. Firstly, alignment errors occur when evaluating detection performance using euclidean distance (see \supref{sec:p_r_f}). For large transients or events containing multiple subdomains, the center identified by an algorithm may be sufficiently displaced from the human-annotated center that the two detections are no longer considered the same event.Detections misclassified due to alignment error are manually marked as TP for the performance evaluation of each algorithm. Examples of this type of alignment error are illustrated in \supfigureref{fig:supp_figure_reverify_examples}A. Secondly, small or dim transients may be missed during human annotation, but picked up by an algorithm. These transients are later marked by a second round of human inspection as TP. Examples of these smaller transients are shown in \supfigureref{fig:supp_figure_reverify_examples}B. A catalogue of revised manual detections for each tested algorithm are included with this publication~\cite{fan_2026_dataset}.

\begin{figure}[htbp!]
\begin{center}
\includegraphics{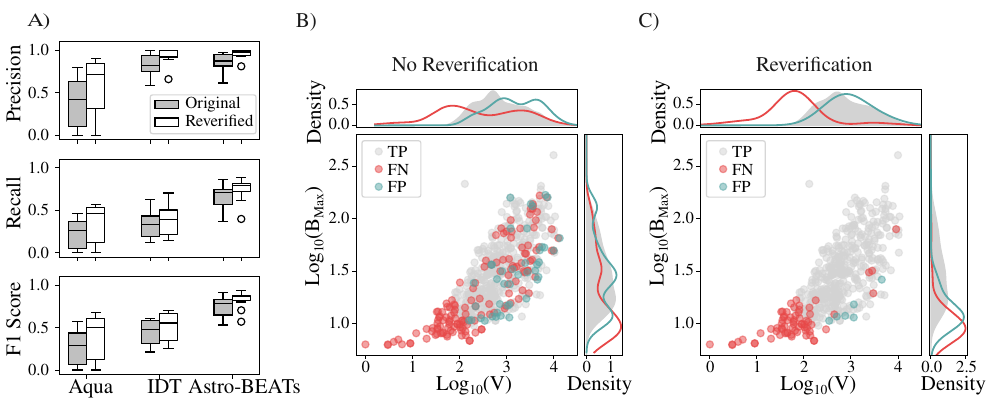}
\end{center}
\caption{\label{fig:supp_figure_reverify} A) Comparison of  the performance of \aqua, \minifinder, and \astrobeats~before and after manual expert reverification. B) We see that, prior to manual expert reverification, many TP detections made by \astrobeats~are falsely labeled as FP or FN, despite their high brightness and volume due to misalignment error. C) Reverification allows us to better quantify the true number of FN and FP detections.}
\addtocounter{suppfig}{1}
\end{figure}

Manual re-verification is only conducted to gain a more accurate measurement of algorithm performance for this work. Comparisons of algorithm performance to manual annotations before and after verification are shown in \supfigureref{fig:supp_figure_reverify}. We present re-verified results throughout this paper for all algorithms in order to give the most accurate assessment of their detection Precision and Recall. We note however, that this second manual re-verification will in general not be possible when \astrobeats~is applied to larger datasets for which we do not have manual annotations.

\section{Detection Metrics} \label{sec:p_r_f}
We evaluate the performance of each algorithm by computing their values of Precision, Recall, F$_1$-score, and DiCE similarity score: \begin{eqnarray}
\mathrm{\textbf{Annotation:}}\nonumber\\
\mathrm{Precision}&=&{\frac {\mathrm{TP}}{\mathrm{TP}+\mathrm{FP}}}
\label{eq:precision} \\
\mathrm{Recall}&=&{\frac {\mathrm{TP}}{\mathrm{TP}+\mathrm{FN}}}
\label{eq:recall}\\
\mathrm{F}_{1}&=&{\frac {2 \times \mathrm{Precision} \times \mathrm{Recall}}{\mathrm{Precision}+\mathrm{Recall}}}
\label{eq:F_beta}\\
\mathrm{\textbf{Segmentation:}}\nonumber\\
\mathrm{DiCE}&=&{\frac {2\times \mathrm{TP}_\mathrm{pix}}{2\times \mathrm{TP}_\mathrm{pix}+\mathrm{FP}_\mathrm{pix}+\mathrm{FN}_\mathrm{pix}}},
\label{eq:dice}
\end{eqnarray}  To assess annotation performance, we calculate a number of TP, FP and FN events for each algorithm using the re-verified dataset. Annotations are matched to manually defined transient centers using the Hungarian algorithm~\cite{kuhn1955hungarian} and the \texttt{linear\_sum\_assignment} function from the Python library \texttt{Scipy}. In this implementation, the Hungarian algorithm assigns centroids detected by each algorithm to manually annotated transient centers with the cost of assignment being the Euclidean distance between the centroids~\cite{beaupre2024quantitative}. We set a maximum allowable separation of 6 pixels in space or 6 frames in time between a centroid identified by the tested algorithm and the corresponding ground-truth centroid for them to be considered a valid match.

To assess segmentation quality, we compare algorithm segmentation masks to those generated by two experts\cite{beaupre2024quantitative} using the Diverse Counterfactual Explanations (DiCE)\cite{dice} similarity score, which evaluates the pixel-by-pixel similarity between the masks generated by a given algorithm and the ground truth annotations. TP$_\mathrm{pix}$, FP$_\mathrm{pix}$, and FN$_\mathrm{pix}$ denote the number of TP (overlapping), FP (algorithm only), and FN (ground truth only) pixels. The DiCE value ranges from 0 (no overlap between ground truth and algorithmic segmentation) to 1 (perfect overlap).

\section{Statistical analysis} \label{sec:statistical tests}
We employ the Student's $t$-test and F-test to assess whether the central tendency ($t$-test) or variance (F-test) of one algorithm differs significantly from another. A difference in model performance is considered statistically significant when the corresponding $p$-value is less than 0.05. In cases where data is not normally distributed, we use the Mann-Whitney test\cite{man1947test} to compare the differences between two groups of observations. To test whether or not the distributions are normally distributed, we use the Shapiro-Wilk statistical test\cite{shapiro/wilk}. We do not report the results of this test, as it was performed solely to determine whether a $t$-test or a Mann–Whitney test was appropriate for comparing the groups.
\clearpage
\section{Supplementary Figures}
\begin{figure}[htbp!]
\begin{center}
\includegraphics[width=0.8\textwidth]{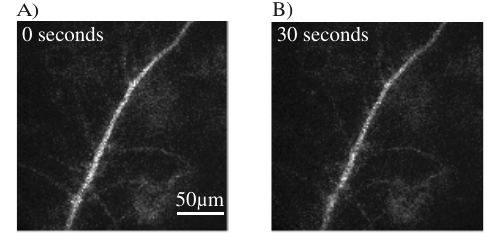}
\end{center}
\caption
{\label{fig:supp_figure_example_fluctuation}Large-scale fluctuation in the brightness of GCaMP6f on the dendritic shaft. The dendrite appears brighter at $t=0~\mathrm{s}$ compared to $t=30~\mathrm{s}$. These brightness fluctuations are accounted for in \astrobeats\ by taking a separate time-dependent background estimate for each frame.}
\addtocounter{suppfig}{1}
\end{figure}

\begin{figure}[htbp!]
\begin{center}
\includegraphics[width=0.8\textwidth,trim={0.2cm 0.1cm 1.35cm 0.1cm}, clip]{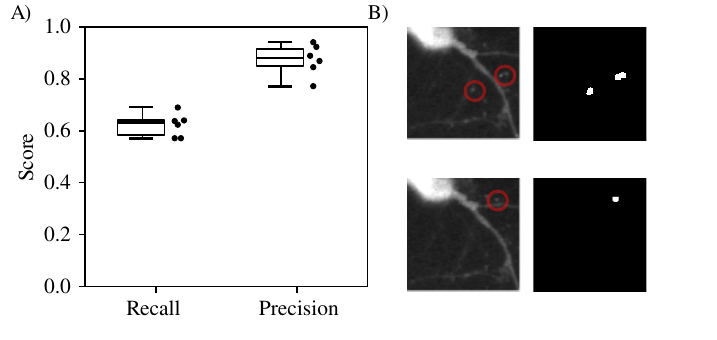}
\end{center}
\caption
{\label{fig:supp_figure_newset} A) Precision and Recall of \astrobeats~ on the out-of-distribution dataset described in Ref.~\citenum{beaupre2024quantitative}. \astrobeats~parameters were not re-optimized for this new dataset and parameters listed in \suptableref{table:dbscan_params} were used. This represents a hypothetical use case for applying \astrobeats~to a new dataset without further fine tuning. B) Example raw data and transient segmentation maps taken from a Ca$^{2+}$-imaging videos of this out-of-distribution dataset. Red circles in the raw video indicate detected transients in the segmentation map.}
\addtocounter{suppfig}{1}
\end{figure}

\clearpage
\section{Supplementary Tables}

\begin{table}[htbp!]
\caption{Parameters of the DBSCAN algorithm. Values used in this work are obtained using the procedure outlined in \supref{sec:optimize} (see \supfigureref{fig:supp_figure_optimization}). \label{table:dbscan_params}} 
\begin{center}    
\begin{tabular}{ |c|c|c|c| }
\hline
\rule[-1ex]{0pt}{3.5ex} Parameter & Description  & Range Tested& Value used here\\ 
\hline\hline
\rule[-1ex]{0pt}{3.5ex} $\mu$ & Initial brightness threshold  & $[3,11]~\sigma$& $7\sigma$\\
\hline
\rule[-1ex]{0pt}{3.5ex} $\mathrm{Min}_\mathrm{pts}$ & Minimum points per cluster & $[100,1000]$& 500\\
\hline
\rule[-1ex]{0pt}{3.5ex} $\epsilon$ & Maximum spatial separation  & $[3,11]$ pixels& 7 pixels\\
\hline
\rule[-1ex]{0pt}{3.5ex} $\epsilon_{t}$ & Maximum time separation  & $[1,3] $ frames& 2 Frames\\
\hline
\end{tabular}
\end{center}
\addtocounter{supptab}{1}
\end{table}

\vspace{0.6cm}
\begin{table}[htbp!]
\caption{Properties of mSCTs calculated for each segmented event in the Ca$^{2+}$-imaging video.}
\label{table:aperture properties}
\begin{center}
\begin{tabular}{ |c|c| } 
\hline
\rule[-1ex]{0pt}{3.5ex} Parameter & Description \\ 
\hline\hline
\rule[-1ex]{0pt}{3.5ex} $A$ &
Area of 2D projection of detection within aperture\\ &
in number of square pixels \\ &
(1px$^2$ = 0.0256$\mu$m$^2$). \\ &
Note that detections with footprint area larger \\ &
than our aperture (\autoref{sec:astrobeats_segmentation}) will be\\ & 
recorded  as filling our entire aperture, having \\ &
an area of 1024 pixels$^2$.\\
\hline
\rule[-1ex]{0pt}{3.5ex} $V$ &
The volume of the detection in number of voxels \\ & 
with  signal above 6$\sigma$ within aperture \\ &
(1 voxel = 0.00256$\mu$m$^2$ms). \\ &
We choose to consider voxels with brightness up \\ &
to 1$\sigma$  below our DBSCAN consideration \\ &
threshold $\mu$, since low brightness FN transients \\ &
may not have pixels above $\mu$.\\ 
\hline
\rule[-1ex]{0pt}{3.5ex} $\langle B\rangle$ & Mean brightness within aperture in $\sigma$\\
\hline
\rule[-1ex]{0pt}{3.5ex}  $\mathrm{B}_{\mathrm{RMS}}$ & Root-Mean-Square brightness  of non-detection pixels in $\sigma$\\
\hline
\rule[-1ex]{0pt}{3.5ex}  $\mathrm{B}_{\mathrm{Max}}$ & Maximum brightness of any voxel in\\
 & background-subtracted aperture in $\sigma$.\\
\hline
\end{tabular}
\end{center}
\addtocounter{supptab}{1}
\end{table}

\vspace{0.6cm}

\begin{table}[ht]
\caption{Range of \aqua~parameters used to optimize algorithm performance.
\label{table:aqua}}
\begin{center}
\begin{tabular}{ |c|c|c| }
\hline
\rule[-1ex]{0pt}{3.5ex} Parameter & Range Tested & Optimal Value Found\\ 
\hline
\rule[-1ex]{0pt}{3.5ex} Smoothing & $[0.5,2]\sigma$ & 0.5$~\sigma$\\
\hline
\rule[-1ex]{0pt}{3.5ex} Minimum size of groups of &&\\ 
connected tentative active voxels & $[5,15]$ voxels & 10 voxels\\
\hline
\end{tabular}
\addtocounter{supptab}{1}
\end{center}
\end{table}

\end{appendices}

\end{spacing}
\end{document}